# Upgrading pulse detection with time shift properties using wavelets and Support Vector Machines

**Jaime Gómez[1], Ignacio Melgar[2] and Juan Seijas[3].**

**Sener Ingeniería y Sistemas, S.A.** [1 2 3]
**Escuela Politécnica Superior, Universidad Autónoma de Madrid. Spain.**[1]
**Departamento de Señales, Sistemas y Radiocomunicaciones, Universidad Politécnica de Madrid. Spain.** [2 3]

jaime.gomez@ii.uam.es
ignacio.melgar@sener.es
seijas@gc.ssr.upm.es

## ABSTRACT

Current approaches in pulse detection use domain transformations so as to concentrate frequency related information that can be distinguishable from noise. In real cases we do not know when the pulse will begin, so we need a time search process in which time windows are scheduled and analysed. Each window can contain the pulsed signal (either complete or incomplete) and / or noise. In this paper a simple search process will be introduced, allowing the algorithm to process more information, upgrading the capabilities in terms of probability of detection ($P_d$) and probability of false alarm ($P_{fa}$).

## 1. INTRODUCTION

In previous work ( [5] and [3] ) we presented a new algorithm for pulse detection using a window where either a complete pulse or noise alone is found. After the input data is sampled, its discrete wavelet transform is computed to obtain wavelet detail coefficients for different scales (see [2], [4] and [6] ). This representation is then introduced into an optimum linear function (a Support Vector Machine) using all available information to make the decision about the presence or absence of such a pulse (see figure 1). This algorithm had a process gain 15 dB better than previous work using wavelet filters alone, i.e., without an optimum information processing decision function, described in [7] and [8].

Note that the output of the linear SVM ( [1] and [9] ) can be either "hard" (if it has two possible values only, 1 and –1) or smooth (when it is a floating point unconstrained number).

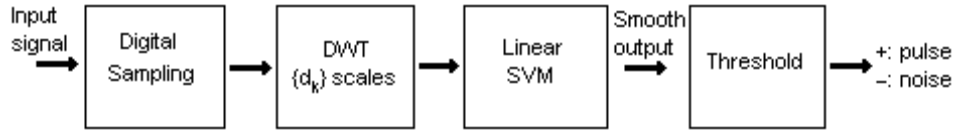

*Figure 1. Pulse detection scheme*

Usually an SVM gives a hard output, but here we have separated the last step: if the smooth output is greater than some threshold the data is positive-class, otherwise it is negative-class. This step loses information, which can be used for further processing, as, for example, class probability. Only in case the decision function is final should the threshold be applied.

But in order to make the algorithm useful in the real world, this approach requires a mechanism to inspect the time domain such that whenever a pulse is emitted you can guarantee that at least one window will be filled with the complete pulse alone.

## 2. TIME-SEARCH PROCEDURE

Our time-search procedure is defined as follows. Given some time interval $[t_0, t_f]$, where amplitude versus time function is defined. Let $T_s$ be the time elapsed between two samples (usually defined by the discretization hardware resources). Let $S$ be the number of samples needed by one complete known pulse. Let $S_t$ be the number of samples (of size $T_s$) between $t_0$ and $t_f$. See figure 2.

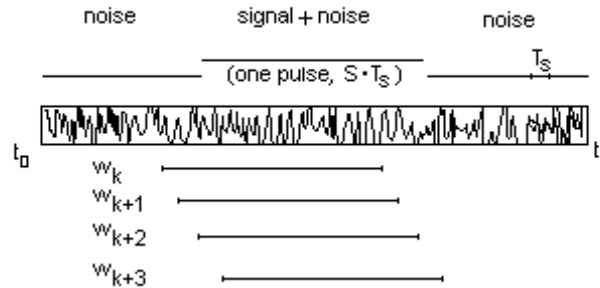

*Figure 2. Data windows in time-search procedure*

Because of the definition of our approach described in [5], windows W must be of the same size as one complete pulse, that is, each window contains S consecutive samples. On interval $[t_0, t_f]$ we will obtain $S_t - S + 1$ windows, such that each window will be defined as the interval $[S_k, S_{k+S}]$. Therefore, two consecutive windows will have the same data except for one sample, and all common samples in the second window will be shifted one position to the left with respect to the first window.

## 3. OUTPUT ANALYSIS

One of the main observations we did was how the window had to fit precisely on the pulse so as to let the SVM perform a correct positive classification. If we execute the complete-pulse optimum SVM over all windows we can see that only a very small subset (less than 7 elements) around the complete-pulse window will show some positive classification degree. All other time-windows behave much as a noise-alone window would about the positive/negative classification rate, regardless of the pulse energy. Therefore, given one single window, we define positive class when at least 99% of the pulse is inside the window, and negative class otherwise.

On the other hand, we can train more SVMs to detect different incomplete pulses, that is, windows were only a portion of the pulse is present plus some noise (to fill the size of the window). Obviously, each one of these SVMs alone will produce poorer results than the complete-pulse SVM, as the detection procedure is applied on less information. Nevertheless, each one processes a different projection (through wavelet transform) of the pulsed signal, and so, they bear different information, as the experiments show. In figure 3 a three-fold multiple window processing is shown. For each window three linear detectors are executed to detect different portions of the pulse. For one pulse, three detectors in three different known windows should detect this event.

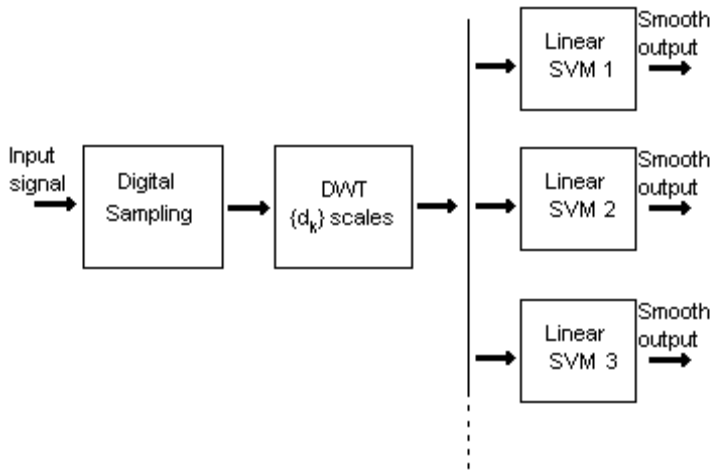

*Figure 3. Multiple SVM window processing*

## 4. EXPERIMENTS

Our experiments had the following setup: chirp pulse (see [3]), 1024 samples size; mother wavelet Daubechies 5, using $d_4$ wavelet coefficients; white Gaussian noise with zero mean and deviation equals one. Three linear detectors (SVMs) were trained against noise alone, such as to detect the complete pulse (named 0-shift), 11 noise samples plus 1013 pulse samples (named 11-shift), and a similar approach for 23 noise samples (named 23-shift) and having $P_{fa} = 10^{-3}$, as established in [3]. Obviously, incomplete-pulse classifiers have less information and will have worse classification rates than its complete-pulse counterpart, but it must still be considered as a source of discriminatory information when used on its own.
For each generated pulse observation, we computed 3 similar Wavelet + Decision-function schemes applied to the complete pulse, incomplete 11-samples-shift pulse, and incomplete 23-samples-shift pulse respectively. For each one, the corresponding above-defined SVM was executed. In figure 2 we described how one processing window contains a subset of sampled input data. In that example, subset $w_{k+2}$ would be defined as the complete-pulse window, and subsets $w_{k+1}$ and $w_k$ would be defined as n-samples-shift pulse, having the first one lower n value than the second one.
Using the three variables as previously defined, the covariance matrix is calculated. Table 1 shows the covariance matrix using 5000 observations on positive data, while table 2 shows similar features for noise-alone observations. As wavelets are linear filters, these matrices shall have equal coefficients using an infinite number of observations, regardless of the signal-to-noise ratio on the pulse or even the presence of signal.

$$\begin{pmatrix} 4.5178 & 2.0537 & 2.9604 \\ 2.0537 & 4.3365 & 1.7090 \\ 2.9604 & 1.7090 & 4.3934 \end{pmatrix}$$

Table 1. Covariance matrix of 5000 pulse observations and three variables (0-shift, 11-shift and 23-shift) and SNR= –15 dB.

$$\begin{pmatrix} 4.5505 & 2.0453 & 3.0958 \\ 2.0453 & 4.3354 & 1.7982 \\ 3.0958 & 1.7982 & 4.6465 \end{pmatrix}$$

Table 2. Covariance matrix of 5000 noise observations and three variables (0-shift, 11-shift and 23-shift).

Note that although the three variables are clearly not independent, they are not completely correlated, that is, all three variables hold some small portion of individual, unshared information. Therefore, as we are able to process this kind of data together efficiently using machine learning algorithms, we can be sure to obtain better results than any of the three sources alone. Moreover, the more incomplete pulse information sources we use, the better results we will obtain. As more variables are introduced in the model, new uncorrelated information will be harder to find, but nevertheless there will always be some degree of upgrade.

## 5. CONCLUSIONS

In this paper we have introduced a new time-search algorithm to allow the linear detector (described in previous papers) to work under real world conditions. This algorithm solves the detector constraints, but it also upgrades the whole system results. In previous work we saw how the application of machine learning models was limitless on the known signal pulse detection problem using wavelets. In this paper we have seen a new unlimited line of information processing: the use of additional Wavelet + Decision-function scheme applied to determined incomplete pulsed signals. This new set of features provides the final model with uncorrelated information, upgrading the classification rates.

Our next step will be to generate a multiple source model using machine learning algorithms (most probably SVMs), to confirm the hints provided by the experiments shown in this paper.

Further analysis is needed to determine how an increased number of processing units in one multiple-source decision function will upgrade probability of detection and / or probability of false alarm.

## 6. ACKNOWLEDGEMENTS

This project is funded by Sener Ingeniería y Sistemas, in the frame of the Aerospace Division R&D program.

## 7. REFERECES

[1] Burges C. "A Tutorial on Support Vector Machines for Pattern Recognition". Knowledge Discovery and Data Mining, 2(2), pp 121-167, 1998.
[2] Daubechies I., "Orthonormal bases of compactly supported wavelets", Communs Pure Appl. Math., 41, 909–996, 1988.
[3] Gomez J., Melgar I., Seijas J., Andina D., "Sub-optimum Signal Linear Detector Using Wavelets and Support Vector Machines", World Scientific and Engineer Academy and Society Conference on Automation and Information (ICAI'03), Tenerife, Spain, December 2003.
[4] Mallat, S., "A theory for multiresolution signal decomposition: the wavelet representation", IEEE Trans. Pattern Analysis and Machine Intelligence, 11, 674–693, 1989.
[5] Melgar I., Gomez J., Seijas J., "Optimum Signal Linear Detector in the Discrete Wavelet Transform – Domain", World Scientific and Engineer Academy and Society Conference on Signal Processing, Computational Geometry, and Artificial Vision (ISCGAV'03), Rhodes, Greece, November 2003.
[6] Palavajjhala, "Computational Aspect of Wavelets and Wavelet Tranforms," Wavelet Application in Chemical Engineering, 1994.

[7] Torres J. Cabiscol P. Grau J., "Radar chirp detection through wavelet transform", Proc. Of the World Automation Congress, WAC2002, Orlando, FL, USA, June 2002.

[8] Torres J., Vega A., Torres S., Andina D., "Chirp Detection Through Discrete Wavelet Transform", Proceedings of the World Scientific and Engineering Academy and Society Conference on Signal Processing, Robotics And Automation (ISPRA'02), 1971-1975, June,2002.

[9] Vapnik V. "The Nature of Statistical Learning Theory". Springer-Verlag, New York, 1995.